# EVALUATING CHUNKING STRATEGIES FOR RETRIEVAL-AUGMENTED GENERATION IN OIL AND GAS ENTERPRISE DOCUMENTS

Samuel Taiwo and Mohd Amaluddin Yusoff

Digital and Innovation Department, Nigeria LNG Limited, Port-Harcourt, Nigeria

## ABSTRACT

*Retrieval-Augmented Generation (RAG) has emerged as a framework to address the constraints of Large Language Models (LLMs), yet its effectiveness fundamentally hinges on document chunking—an often-overlooked determinant of its quality. This paper presents an empirical study quantifying performance differences across four chunking strategies: fixed-size sliding window, recursive, breakpoint-based semantic, and structure-aware. We evaluated these methods using a proprietary corpus of oil and gas enterprise documents, including text-heavy manuals, table-heavy specifications, and piping and instrumentation diagrams (P&IDs). Our findings show that structure-aware chunking yields higher overall retrieval effectiveness, particularly in top-K metrics, and incurs significantly lower computational costs than semantic or baseline strategies. Crucially, all four methods demonstrated limited effectiveness on P&IDs, underscoring a core limitation of purely text-based RAG within visually and spatially encoded documents. We conclude that while explicit structure preservation is essential for specialised domains, future work must integrate multimodal models to overcome current limitations.*

## KEYWORDS

*RAG, AI, Oil and Gas, Information Retrieval*

## 1. INTRODUCTION

The rise of Generative Artificial Intelligence (GenAI) popularised by Chat Generative Pre-Trained Transformer (ChatGPT) – an instruction-fine-tuned large language model (LLM) developed by OpenAI - has led to an increasing interest in various sectors for small and medium-sized enterprises to incorporate LLMs into diverse facets of their business. [15], [17].

Despite LLMs' advanced understanding of natural language, they remain constrained by their static pretraining on a fixed corpus of knowledge – usually general and publicly available information – and a very limited context window [11]. A significant limitation of LLMs is the tendency to generate "hallucination" or create fabricated information [5] with full confidence. Additionally, the lack of specific proprietary knowledge, limited context window, data freshness, and factual inaccuracy continues to plague LLMs; particularly relevant in knowledge-intensive tasks like open-domain question answering [13], [23], [15], [9].

In 2020, Lewis P. et. al. [10] introduced RAG, a novel technological paradigm to address the aforementioned challenges faced by LLMs, by enhancing LLMs with external knowledge [10]. RAG has rapidly become the de facto for an advanced zero-shot information retrieval solution for documents [15] since its introduction in 2020. Diverse research on optimisation techniques, variants of RAG, and evaluation techniques has been developed, focusing mainly on two aspects: improving retrieval accuracy [19] and enhancing the robustness of LLMs against toxic information [8]. However, RAG's ability to perform well fundamentally relies on high-quality document chunking – the method used to segment documents into digestible and accessible parts [9] – a commonly overlooked aspect [22]. Traditional text chunking methods rely on simple text extraction, often based on rules, semantic similarities, or fixed-length chunks, which compromise and fail to preserve the structural integrity of complex documents [20], [18].

Fixed-length document segmentation, as demonstrated by Boheng et al. [18], frequently creates breakpoints mid-sentence. This method ensures that only a small fraction of sentences is fully preserved within any single chunk [18]. Consequently, when sentences containing crucial information or answers are fragmented, their meaning is at risk of being distorted, which can lead to the exclusion of necessary related sentences and, ultimately, inaccurate responses [20]. This negative cycle initiates a "weakest link effect," where the initial quality of text chunking imposes a severe constraint on the retrieved content, thus limiting the accuracy of the final generated output [17], [23].

Globally, the oil and gas (O&G) industry grapples with an immense repository of technical documentation, including contracts, regulatory compliance records, geological reports, and detailed engineering schematics. This massive volume creates a critical operational bottleneck, with estimates suggesting that up to 80% of employee time is consumed navigating and searching through structured and unstructured data to inform mission-critical decisions [16]. Operational hurdles include limited accessibility, severe version control issues, and inefficient retrieval processes across diverse geographic locations [12].

The intrinsic complexity of O&G documentation, such as Piping and Instrumentation Diagrams (P&IDs) and specification reports, necessitates specialised handling. These documents are often multimodal, combining dense text with highly structured data, engineering symbols, diagrams, and tables [2]. Unlike linear, homogeneous text, these documents rely on visual layout and structural hierarchy to convey meaning. For instance, a safety procedure may span multiple pages, with critical numerical data contained in a visually distinct table adjacent to explanatory text.

In response to the aforementioned challenges of traditional chunking strategies and the heterogeneous structure of oil and gas documents, this paper outlines a comprehensive empirical study designed to quantify the performance differential among four distinct chunking strategies:
1.   Fixed-size sliding window
2.   Recursive (hierarchical splitting)
3.   Breakpoint-based Semantic
4.   Structure-aware.

When applied to a corpus of complex, structured oil and gas enterprise documents that represent the following types:
1.   Text-heavy with structural layout (e.g., policies, procedures, safety manuals)
2.   Table-heavy (e.g., equipment specs)
3.   Diagram-referencing (piping and instrumentation diagrams)

## 2. RELATED WORK

Since the emergence of using RAG to augment the capabilities of LLMs, there has been an ample amount of research focused on diverse chunking strategies, but little to no research has been carried out to evaluate how different chunking strategies affect retrieval performance, especially when considering the complex structure of oil and gas documents.

## 2.1. Structured Data

Enterprise data typically consists of a combination of formats, including structured, semi-structured, and tabular information. These data are often encapsulated in structured records, unstructured records, policy documents, and tabular formats that LLMs can't directly process or reason over [3]. This challenge is further heightened in industries like oil and gas, where essential data, such as P&IDs and isometric drawings, combines both unstructured records and tabular components. Therefore, preprocessing is necessary before LLMs can effectively ingest and utilise this type of enterprise data.

The emergence of RAG has addressed the gap in integrating such document formats with LLMs, enabling efficient information retrieval [10]. However, as observed by Chandana et. al. [3], there are some key limitations when using baseline RAG methods optimised primarily for unstructured data. Some of these limitations include:

1. *Fragmented Contextual Representation*: The use of basic, fixed token-length splits for chunking often leads to fragmentation of essential contextual information. This issue is especially problematic when dealing with intricate documents such as legal policies or user manuals [3], [6].
2. *Inadequate Handling of Tabular Data*: Flattening tabular data (tables) into a continuous text stream eliminates the essential row-column structure, which is necessary for precise retrieval of information contained within those tables [3].

## 2.2. Chunking Strategies

The success of a RAG workflow remains tied to the chunking strategy used to segment the document. A good chunking strategy can lead to a well-optimised retrieval quality and computational efficiency [4].

### 2.2.1. Fixed-sized chunking

Fixed-size chunking is used as our baseline chunking that splits the document into a fixed-sized chunk on a predefined chunk size. It is one of the simplest methods for segmenting text into smaller chunks, as this approach is simple and computationally efficient. We also employ the use of sentence overlap between chunks to maintain a sense of continuity, as using just chunk size usually leads to a potential retrieval degradation, as there will be no specific understanding of the document [14].

### 2.2.2. Recursive Chunking

Recursive chunking addressed some of the limitations of fixed-size chunking and is usually the go-to baseline for most RAG applications. It operates by dividing the text hierarchically using a set of separators to split the text. If the output chunks are larger than the desired size, it recursively applies the next separator in the hierarchy [21].

### 2.2.3. Breakpoint-based Semantic Chunking

Breakpoint-based semantic chunking refines chunking by analysing the sequence of sentences to determine optimal breakpoints for splitting the text into two segments. A breakpoint is inserted when the semantic distance between two sentences is higher than a set threshold, signalling a significant topic change [14].

### 2.2.4. Structure Aware Chunking

This is a simple yet effective method that follows the document structure when chunking. It operates by dividing the text hierarchically using a set of headers to split based on. This is similar to recursive chunking, but instead of text separators, header separators are used.

## 2.3. Comparison with Existing Approaches

While existing chunking methods have their strengths and have laid a robust foundation for this work, they often fail to capture the challenges posed by documents with heterogeneous data formats. This paper highlights that for enterprise heterogeneous documents, more importantly, oil and gas documents with a combination of multiple data formats and isometric diagrams, using structure-aware chunking when processing these documents for RAG improves the retrieval system, thereby eliminating the "weakest link effect" and improving the overall RAG workflow.

## 3. METHODOLOGY AND EXPERIMENTAL SETUP

To guide our study, we formulated three research questions (RQs), with the intention of directing the research to a more concrete, evidence-based approach. The following RQs guide the experimental design and analysis of this study:

**RQ1:** How does structure-aware chunking compare against semantic chunking as well as established baselines (Fixed-Size and Recursive Text Splitting) in terms of overall retrieval effectiveness and the quality of retrieved context?
**RQ2:** Can structure-aware chunking achieve retrieval performance comparable to semantic chunking while demonstrating significant improvements in indexing efficiency, storage requirements, and computational latency?
**RQ3:** What is the retrieval performance limitation across all four chunking strategies when applied to documents with high visual and spatial complexity, specifically the P&IDs?

Due to the sensitive nature of the data and adherence to enterprise security protocols, all experiments are strictly confined to the designated cloud environment, prohibiting the use of unvetted, external open-source models.

## 3.1. Document Preprocessing

The underlying knowledge base for all the experiments carried out was a collection of sixty proprietary documents. This was done to ensure that the dataset reflects a comprehensive and unbiased sample of the challenges faced in a real-world O&G environment. The corpus was stratified into three distinct document types, with twenty documents randomly selected from each category:

1. *Text-heavy Documents:* Documents primarily composed of contiguous narrative text, such as HSE (Health, Safety, and Environment) safety operating procedures (SOCs), policy documents, and administrative guidelines. These documents test the system's ability to handle long-form, context-dependent text.

2. *Table-heavy Documents:* Documents where information is predominantly presented in structured tables, such as field design change, compliance reports, and regulatory checklists. These test the indexing system's resilience to mixed text/data formats.
3. *Piping and Instrumentation Diagrams Documents:* Documents containing significant visual and spatial information (e.g., engineering diagrams) where long text is sparse, and includes highly structured tables and isometric information. These documents introduce challenges related to spatial reasoning and the ability of chunking strategies to capture minimal, yet crucial, descriptive text.

Documents were extracted using Azure Document Intelligence (ADI). The selection of ADI was based on its advanced capabilities for structural preservation and metadata capture, which are essential for enhancing downstream RAG performance, which is also an advantage for structure-aware chunking.

The primary advantages leveraged from ADI for this methodology include:

1. *Structure-Preserving Extraction:* ADI performs a non-linear, structure-preserving extraction of tabular content. Tables are represented in a structured markdown format, which ensures the relational integrity (rows, columns, headers) of the original data is maintained, facilitating structurally aware chunking.
2. *Structured Output for Chunking:* ADI provides the document text in a Markdown format, which intrinsically encodes structural information (headings, lists, tables). This output is directly beneficial for chunking strategies that utilise the document's inherent hierarchy, specifically, structure-aware chunking.
3. *Comprehensive Metadata Capture:* The extraction process captures critical metadata necessary for retrieval and post-processing. This includes bounding box coordinates (text location), logical document sections, figures, and other structural elements.

Using Azure Document Intelligence ensures robust text and table extraction, ready for chunking.

## 3.2. Chunking Strategies

As stated above, four different chunking strategies were used, and fixed-sized chunking served as our baseline.

*Fixed-Sized Chunking.* Documents were split into fixed-sized chunks of 512 characters with a character overlap of 128. This method serves as a simple baseline, prioritising computational speed over semantic coherence.

*Recursive Chunking.* This strategy recursively splits the text using a sequence of increasingly smaller delimiters of ["\n\n", "\n", " ", ""]. This attempt to keep paragraphs and sections intact thereby preserving local content. For each recursion, we used a chunk size of 1024 and a chunk overlap of 100.

*Breakpoint-Based Semantic Chunking.* Leveraging an embedding model, we used a breakpoint-based semantic chunking with a percentile threshold of 0.8 to identify and group text segments that are semantically coherent above the specified breakpoint.

*Struct-Aware Chunking.* We utilised the inherent structure of the document to define boundaries and breakpoints while chunking. We used a header-based splitting method to separate the text into three header types – H1, H2, and H3.

Each chunk produced by the four different methods was indexed separately, creating four parallel document stores to be queried during the polling phase.

### 3.3. Embedding and Indexing

We used OpenAI text-embedding-3-large for embedding the chunks, which were then indexed into an Azure CosmosDB vCore vector database with a Hierarchical Navigable Small World (HNSW) index type. We used four different indices for the four chunking strategies. This is to ensure efficient analysis for each chunking strategy.

### 3.4. Retrieval Strategy

Dense Retrieval (Vector Search):	The chunks were transformed into a high-dimensional numerical space representation using a text-embedding-3-large embedding model [1]. These embeddings were then stored in an HNSW index.

### 3.5. Dataset Creation and Annotation

As there were no publicly available datasets that met the diverse and domain-specific requirements necessary for evaluating an information retrieval system, specifically within the oil and gas industry, we created our training and evaluation dataset. The underlying knowledge base for the experiments was a collection of sixty proprietary documents.

#### 3.5.1. Query Generation and Run Pooling

To ensure that the resulting ground truth reflected realistic information needs and covered a diverse range of documents in the corpus, we followed the standard pooling methodology used in Information Retrieval (IR) test collections, specifically the Text REtrieval Conference (TREC) format.

A set of Q evaluation queries (where Q is 60) was created. These queries were designed to test the different facets of the retrieval system. These queries were generated semi-automatically using LLM with a human in the loop, ensuring that the queries are up to par and are not biased towards a section of the document.

The pooling technique was used to select manageable documents for human annotation, maximising the probability that relevant documents are included. The pooling process was as follows:

1. *System Runs:* The Q queries were run against the primary retrieval systems across all four chunking strategies.
2. *Pool Creation:* For each query, the top K (where K is 10) was retrieved by each of the distinct system runs, and these were combined into a single, merged document pool. Duplicate documents were removed, and the remaining unique document-query pairs formed the candidate set for human judgment.

#### 3.5.2. Manual Relevance Annotation

The pooled documents-query pairs were subject to manual relevance by human annotators using the Label Studio annotation tool.

Rather than using binary assessment (Relevant/Not Relevant), we employed a three-point graded relevance scale to capture the nuances in document, utility, which is essential for metrics like Normalised Discounted Cumulative Gain (NDCG):

1. *Grade 0 (Not Relevant)*: The chunk contains no useful information related to the query's intent.
2. *Grade 1 (Partially Relevant)*: The chunk contains some information related to the query, but it is incomplete for the query's intent.
3. *Grade 2 (Highly Relevant)*: The chunk explicitly and fully contains the information necessary to answer the query.

The final product of the annotation process is the Query Relevance Judgments (Qrels) file, which adheres to the standard IR format:

*query_id        0        document_id            score*

The qrels serve as the ground truth for evaluating the retrieval effectiveness of each chunking strategy and how each chunk works well with the different document variety.

## 3.6. Evaluation Metrics

We evaluated the system using standard information retrieval metrics, focusing on both the precision of the top-ranked results and the overall effectiveness across the ranked list. We utilised four key metrics: F1-score, NDCG, MRR, and MAP, where each metric served a specific purpose in assessing different aspects of the results.

### 3.6.1. F1-Score

The F1 score is a harmonic mean of precision, where an F1 score reaches its best value at 1 and the worst score at 0. It is used to balance the importance of precision and recall.

### 3.6.2. Normalised Discounted Cumulative Gain (NDCG)

NDCG is a metric for evaluating the quality of ranking in information retrieval. It evaluates the usefulness of an item based on its position in the rank, assigning higher weights to items appearing at the top of the list. Because our dataset uses a three-point graded relevance scale, NDCG is necessary for ranking highly relevant chunks over partially relevant chunks.

### 3.6.3. Mean Reciprocal Rank (MRR)

MRR is a statistical measure for evaluating the effectiveness of a retrieval system. It focuses on the position of the first relevant item in the ranked list.

### 3.6.4. Mean Absolute Precision (MAP)

MAP is an evaluation metrics that calculates the mean of the average precision score across a set of queries that measures the ranking quality of the retrieved chunks.

## 4. RESULT AND ANALYSES

### 4.1. Structure-Aware Chunking Versus Established Chunking Strategies

The results from Table 1, although nuanced, reveal that structure-aware chunking is significantly more successful at placing the first and most relevant piece of information at the top of the list. NCDG@3 and NCDG@5 confirm that by leveraging the inherent document structure, the most highly relevant chunks are consistently ranked highest, providing the LLM with immediate and high-quality context. This finding validates the core benefits of structure-aware chunking in a specialised domain like oil and gas.

Recursive and fixed-size chunking however achieved the highest score in MAP and F1-score, suggesting that while structure-aware chunking excels at getting the absolute best result, recursive and fixed-sized chunking demonstrate greater consistency in retrieving a larger number of relevant documents across the entire list. This is largely due to the chunk size distribution and potential indexing redundancies.

The most unexpected finding is the underperformance of breakpoint-based semantic chunking, which scored lowest across all high-relevance metrics, suggesting that the strategy's inability to interpret the technical and structural nature of the oil and gas documents is a significant drawback. The generalised nature of the semantic embedding model leads to poor boundary determination and subsequent low relevance scores.

Table 1. Comparative results of all four chunking strategies. Bold values indicate the best performance for each metric. Superscripts a, b, c, and d indicate statistical significance (where $p < 0.05$) over fixed-size (a), recursive (b), semantic (c), and structure-aware (d) chunking strategies, respectively using Fisher's Randomization Test.

| Strategy | MRR | NDCG@3 | NDCG@5 | MAP@3 | MAP@5 | F1@3 | F1@5 |
|---|---|---|---|---|---|---|---|
| fixed-size | 0.575 | 0.366[c] | 0.326[c] | 0.101 | 0.124 | 0.161 | 0.183[cd] |
| recursive | 0.589[c] | 0.398[ac] | 0.353[c] | **0.109** | **0.124** | **0.166[cd]** | **0.184[cd]** |
| semantic | 0.557 | 0.305 | 0.270 | 0.098 | 0.115 | 0.151 | 0.167[d] |
| struct-aware | **0.644[abc]** | **0.409[ac]** | **0.361[ac]** | 0.104 | 0.117 | 0.150 | 0.152 |

## 4.2. Computational Cost of Structure-Aware Chunking Versus Semantic Chunking

The analysis in Table 2 focuses on the computational cost of all four strategies, which are directly proportional to the number of documents processed. Structure-aware chunking performs well in minimising fragmentation, resulting in only 911 chunks and an index size of 15MB, representing just about 8% of the total size of semantic chunking and 30% of both recursive and fixed chunking. This efficiency stems from the strategy's ability to structurally group related text into single, large, high-context chunks, thereby preventing unnecessary fragmentation.

The retrieval speed strongly correlates with the size and dimensionality of the index, as the structural chunking process yields the fastest average retrieval time, compared to semantic chunking, which has the lowest average retrieval time, demonstrating the severe performance penalty incurred within a highly fragmented index.

This result also solidifies structure-aware chunking as a strategy that achieves high-quality context with the lowest operational latency.

Table 2. Comparative computational and index across all chunking strategies.

| Strategy | Total No. of Chunks | Index Storage Size (MB) | Avg. Retrieval Time (MS) |
|---|---|---|---|
| fixed-size | 4014 | 63.789 | 510.4 |
| recursive | 2637 | 41.937 | 567 |
| semantic | 10685 | 169.719 | 846 |
| struct-aware | **911** | **15** | **431** |

## 4.3. Retrieval Limitation on Piping and Instrumentation Documents

The low metrics score shown in Table 3, reflects the inherent difficulty of applying text-based retrieval methods to visually dense engineering documents specifically PIDs [1]. Despite this challenges, structure-aware chunking still poses a significant statistical ability to identify and preserve key blocks of text, significantly outperforming fixed-size, recursive, and semantic chunking strategies.

Although the limited sample size of P&ID documents restricts the magnitude of achievable scores, the consistent relative advantage of structure-aware chunking suggests that explicitly preserving document layout and structural cues is more effective than purely text-driven segmentation methods for visually complex documents.

These findings align with recent work by Achmad et al. [1], who demonstrated the potential of graph-based representations combined with large language models for intuitive retrieval and analysis of P&IDs. While such approaches show promise, the results in Table 3 indicate that challenges remain in improving retrieval accuracy for PIDs, highlighting the need for hybrid representations that integrate textual structural, and visual information.

Table 3. Comparative results of all four chunking strategies across visually complex documents
Superscripts a, b, c, and d indicate statistical significance (where $p < 0.05$) over fixed-size (a), recursive (b), semantic (c), and structure-aware (d) chunking strategies, respectively using Fisher's Randomization Test.

| Strategy | MRR | NDCG@3 | NDCG@5 | MAP@3 | MAP@5 | F1@3 | F1@5 |
|---|---|---|---|---|---|---|---|
| fixed-size | 0.070[b] | 0.042[b] | 0.039[b] | 0.016[b] | 0.019[b] | 0.025[b] | 0.027[b] |
| recursive | 0.049 | 0.027 | 0.024 | 0.011 | 0.013 | 0.019 | 0.019 |
| semantic | 0.070[b] | 0.039[b] | 0.036[b] | 0.020[b] | 0.021[b] | 0.024 | 0.024 |
| struct-aware | **0.106[abc]** | **0.066[abc]** | **0.057[abc]** | **0.032[abc]** | **0.035[abc]** | **0.040[abc]** | **0.035[abc]** |

## 5. CONCLUSION

The results successfully demonstrated the effectiveness of document chunking in a RAG system on O&G document type. Structure-aware chunking proved to be the most critical discovery, consistently achieving the highest performance in top-K metrics across the overall corpus, validating the core hypothesis that explicit structure preservation is essential in specialised domains like oil and gas. Structure-aware chunking achieved significant efficiency, showing that minimising data fragmentation would result in faster operational latency, which is optimal for a production system where accuracy and computational efficiency are paramount.

However, the results exposed a major universal limitation of the current text-based RAG system struggle with visually complex documents, such as P&IDs diagrams. Although, structure-aware method shows significant statistical ability of effectively preserving the few coherent text blocks it received, the results necessitate a focus on multimodal RAG approaches that integrate structure and visual cues alongside text. Future work must integrate visual layout with multimodal architectures, and refined retrieval granularity to correctly sequence complex text and overcome the fundamental limitations that currently prevent reliable information retrieval from non-standard, visually dense engineering documents.

## AUTHORS

Samuel Taiwo is a machine learning engineer with 6+ years of experience building machine learning system at scale. Currently at Nigeria LNG, he focusses on driving digital transformation through AI-advocation for its implementation across the organisation. He enjoys implementing research work to real-world applications and scaling theoretical approaches.

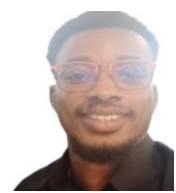

Dr. Mohd Amaluddin Yusoff is an Analytics and AI practitioner with experience across oil and gas, academia, and industrial engineering. He currently works on applying data analytics and artificial intelligence to practical business problems, including predictive maintenance and generative AI. With a background in engineering and applied mathematics, he enjoys bridging theory and real-world applications and is particularly interested in building internal capabilities through mentoring, and knowledge sharing.

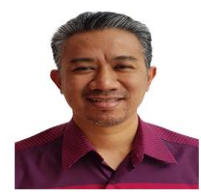